\documentclass[fleqn,usenatbib]{mnras}

\usepackage[T1]{fontenc}

\DeclareRobustCommand{\VAN}[3]{#2}
\let\VANthebibliography\thebibliography
\def\thebibliography{\DeclareRobustCommand{\VAN}[3]{##3}\VANthebibliography}

\usepackage{graphicx}	
\usepackage{amsmath}	

\providecommand{\abs}[1]{\lvert#1\rvert}
\usepackage[dvipsnames]{xcolor}
\newcommand{\roc}[1]{\textcolor{Magenta}{#1}}

\usepackage{newtxtext,newtxmath}

\title[planet migration via planet-vortex interactions]{Spreading pressure bumps in gas-dust discs can stall planet migration via planet-vortex interactions}

\author[R. O. Chametla and O. Chrenko]{
R. O. Chametla$^{1}$\thanks{E-mail:raul@sirrah.troja.mff.cuni.cz (ROC)}
and O. Chrenko$^{1}$\\
$^{1}$Charles University, Faculty of Mathematics and Physics, Astronomical Institute, V Hole$\check{s}$ovi$\check{c}$k\'ach 747/2, 180 00 Prague 8, Czech Republic\\ 
}

\date{Accepted XXX. Received YYY; in original form ZZZ}

\pubyear{2015}

\begin{document}
\label{firstpage}
\pagerange{\pageref{firstpage}--\pageref{lastpage}}
\maketitle

\begin{abstract}
We investigate the gravitational interaction between low- to intermediate-mass planets ($M_p \in[0.06-210]\,M_{\oplus}$) and two previously formed pressure bumps in a gas-dust protoplanetary disc. We explore how the disc structure changes due to planet-induced perturbations and also how the appearance of vortices affects planet migration. 
We use  multifluid 2D hydrodynamical simulations
and the dust is treated in the pressureless-fluid approximation, assuming a single grain size of $5\,\mu{\mathrm{m}}$. The initial surface density profiles containing two bumps are motivated by recent observations of the protoplanetary disc HD163296. When planets are allowed to migrate, either a single planet from the outer pressure maximum or two planets from each pressure maximum, the initial pressure bumps quickly spread and merge into a single bump which is radially wide and has a very low amplitude. The redistribution of the disc material is accompanied by the Rossby Wave Instability (RWI) and an appearance of mini-vortices that merge in a short period of time to form a large vortex. The large vortex induces perturbations with a spiral wave pattern that propagate away from the vortex as density waves. We found that these vortex-induced spiral waves strongly interact with the spiral waves generated by the planet and we called this mechanism the "\textit{Faraway Interaction}". 
It facilitates much slower and/or stagnant migration of the planets and it 
excites their orbital eccentricities in some cases. Our study provides a new explanation for how rocky planets can come to have a slow migration in protoplanetary discs where vortex formation occurs.
\end{abstract}

\begin{keywords}
Planetary systems: protoplanetary discs -- hydrodynamics -- instabilities
\end{keywords}



\section{Introduction}

Recent observations of young protoplanetary discs around stars of intermediate mass using several techniques and different instruments (e.g. the Atacama Large Millimeter/submillimeter Array (ALMA), Very Large Array (VLA), Keck II and Hubble telescopes) have revealed structures such as spiral arms \citep{Garufi2013,Grady2013,Benisty2015,Reggiani2018}, gaps and bright rings \citep{Quanz2013}, large cavities \citep{Andrews2011}, asymmetries \citep{van2013} and point-like sources \citep{Reggiani2018}. The formation of different structures in protoplanetary discs may be the result of various processes: the aerodynamic coupling of gas and dust; the dynamics driven by self-gravity; the hydrodynamic or magnetohydrodynamic turbulence, among others. Many recent studies suggest that the most likely agent responsible for the formation of these structures are interactions of planets with the disc at an early stage of their formation \citep{Kep2018,Muller2018,Pinte2018,Pinte2019}.

It is particularly interesting to consider the shape and time evolution of the observed dark gaps and bright rings. A bright ring is believed to be a radial pressure bump in the gas disc where dust accumulation occurs \roc{\citep[][]{2016PhRvL.117y1101I,2018ApJ...869L..41A,2018ApJ...869L..42H,2018ApJ...869L..49I,2018ApJ...869L..50P,2020ApJ...890L...9P,2021ApJS..257...18T}.} The accumulation is facilitated by a positive radial pressure gradient that induces super-Keplerian gas flow in which dust grains suffer tail winds, while a negative radial pressure gradient induces head winds \citep{Klahr2005}. Several analytical studies and numerical simulations have been carried out to investigate the temporal evolution of these pressure bumps with no planet embedded in the disc \citep{Lovelaceetal1999,Lietal2000,Lietal2001,Meheutetal2010,Takietal2016}. Once a pressure bump is formed it can be maintained for a long time if it meets the stability criteria of Solberg-Hoiland \citep{Lietal2000,Lietal2001}. If the criteria are violated, on the other hand, the pressure bump is destabilized and the Rossby wave instability (RWI) is inevitably triggered \citep{Lovelaceetal1999,Meheutetal2010}. The instability generates Rossby vortices that grow exponentially with time and redistribute the angular momentum of the disc.

The ringed structure of discs of gas and dust can strongly modify the migration of embedded
low-mass planets. For example, a pressure bump can halt inward Type I migration \citep{Tanaka2002,Masset2006}
due to local variations of the vortensity gradient that can make the corotation torque felt by planets
positive. Additionally, recent numerical studies of Super-Earth planets migrating in gas and dust discs show that planets can experience periods of runaway migration as a result of dust redistribution at pressure peaks and gaps \citep{Dong2017,Dong2018,Weber2019,Gaylor2020}. These studies suggest that the pressure bumps and partial gaps are formed by the same planet during its migration.

Here, instead, our main hypothesis is that the planet begins to migrate in the presence of pressure bumps that have already been formed by some unspecified mechanism
(we introduce the bumps regardless of the mechanism that created them and there is also no additional mechanism, beside the initial rotational equilibrium, that would prevent the bumps from spreading). 
We explore how the pressure bumps evolve in the presence of low- to intermediate-mass non-accreting planets that are allowed to migrate freely in the disc. We also investigate how the interaction of planets with vortices, which form due to the redistribution of gas and dust, modifies planetary migration.

Motivated by recent observational advances, our disc model resembles HD163296, which is a system
located at a distance of 101.5 pc \citep[see][]{2018A&A...616A...1G}. It consists of a young Herbig Ae star with the mass $M_\star=2.02\,M_\odot$ \citep[][]{2019Natur.574..378T} and a gas-dust disc with 
an estimated radius of 550 au. The disc contains two well defined rings located at $68$ and $100$ au and a
less sharp third ring at $159$ au that have been observed in the mm-continuum emission \citep[see][]{2018ApJ...869L..49I}. It is believed that the rings are accumulations of mm-sized dust grains trapped within gas pressure maxima \citep[][]{2018ApJ...860L..12T} while $\mu$m-sized dust grains reside in regions of low dust density \citep[][]{2019ApJ...886..103O}, in addition to a depletion of the gas within the dust gaps \citep{2016PhRvL.117y1101I}. However, an adequate modeling of the spatial distribution of the dust in HD163296 has emerged as a complex challenge \citep[see][and references therein]{2018ApJ...857...87L}. Recently \citet[][]{2021ApJ...912..164D} constrained the dust scale height by comparing the DSHARP high-resolution millimeter dust continuum image with radiative transfer simulations. Their results suggest that the gas turbulence is stronger or the dust grains are smaller at the inner ring than at the outer ring.

\begin{figure}
	\includegraphics[scale=0.5]{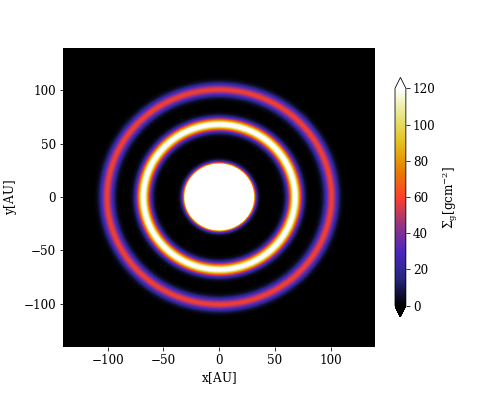}
    \caption{2D map of the initial gas surface density at $t=0$ orbits without any planet embedded in the disc. The dense rings trace the pressure bumps. The radial extent of the disc is from $10$ to $200$ au. The pressure maxima are located at $r_{\mathrm{b1}}=67.9$ au and $r_{\mathrm{b2}}=100.5$ au, respectively.}
    \label{fig:initial}
\end{figure}

In our study, we adopt a disk model with a radial extension and density profiles similar to those reported in \citep[][]{2021ApJ...912..164D}. However, we consider (i) only one size of dust grains
and we assume (ii) that they are strongly aerodynamically coupled to the gas. These simplifications are then used to (i) scale the observed dust density profile directly to a gas density profile and (ii)
ensure a long-term stability of dust concentrations within pressure maxima without introducing
any restoring viscosity transitions.
Our simplifications are justifiable as we are mainly interested in analyzing the type I planetary
migration and not in explaining the observed structures in HD163296. Fig.~\ref{fig:initial}
shows the initial surface densities of gas and dust in the disc model that we consider.
There are two dense rings that are caused by two pressure bumps.

The paper is organized as follows. In Section~\ref{sec:model}, we describe the governing equations of our hydrodynamic model. We also specify the numerical code, initial conditions and parameters. In Section~\ref{sec:results}, we present the results of 2D hydrodynamic simulations and a brief discussion. Finally, the main conclusions are given in Section~\ref{sec:conclusions}.

\section{Physical Model}
\label{sec:model}

We study the evolution of a 2D disc on a uniformly spaced polar Eulerian mesh characterized by the radial coordinate $r$ and the azimuthal angle $\phi$. The disc consists of two fluids representing gas and one dust species.

\subsection{Gas}

Hydrodynamic equations describing the gas flow are the equations of continuity
\begin{equation}
\frac{\partial\Sigma_{\mathrm{g}}}{\partial t}+\nabla\cdot(\Sigma_{\mathrm{g}}\mathbf{u})=0,
\end{equation}
and the gas momentum equation
\begin{equation}
\Sigma_{\mathrm{g}}\frac{D\mathbf{u}}{Dt}=-\nabla P-\nabla\cdot\tau-\Sigma_{\mathrm{g}}\nabla\Phi-\Sigma_{\mathrm{d}}\mathbf{f}_{\mathrm{d}},
\end{equation}
where $\Sigma_{\mathrm{g}}$ is the gas surface density, $\mathbf{u}$ is the gas velocity, $P$ is the pressure, $\tau$ is the viscous stress tensor, $\Phi$ is the gravitational potential, $\Sigma_{\mathrm{d}}$ is the surface density of dust, $\mathbf{f}_{\mathrm{d}}$ is a function that represents the interaction between gas and dust grains via the aerodynamic drag force and the Lagrangian derivative is defined as
\begin{equation}
\frac{D}{Dt}\equiv\frac{\partial}{\partial t}+\mathbf{u}\cdot\mathbf{\nabla}.
\end{equation}

We assume a vertically isothermal disc for which the pressure is given as
\begin{equation}
P=c_{\mathrm{s}}^2\Sigma_{\mathrm{g}},
\label{eq:eos}
\end{equation}
where
\begin{equation}
c_{\mathrm{s}}(r)=\frac{H(r_0)}{r_0}\left(\frac{r}{r_0}\right)^f\Omega_{\mathrm{K}}r,
\end{equation}
is the local isothermal sound speed, $\Omega_{\mathrm{K}}$ is the Keplerian angular frequency and $f=0$ is the flaring index. Here $r_0$ is a reference radius and the aspect ratio $h=H/r$ is taken to be constant through the disc.

\begin{figure*}
	\includegraphics[scale=0.55]{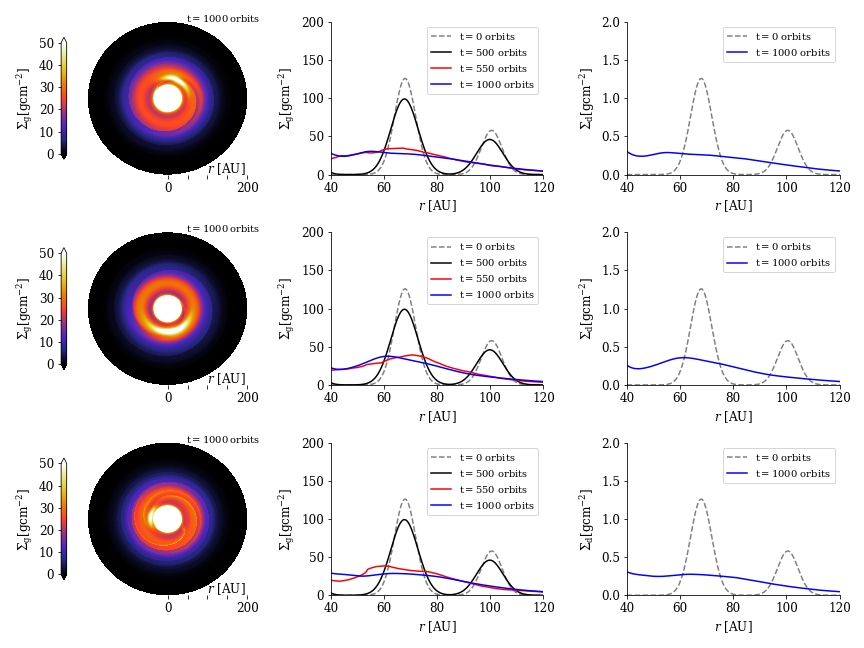}
    \caption{2D map of the gas surface density $\Sigma_{\mathrm{g}}$ (\textit{left column}), temporal evolution of the radial profile of $\Sigma_{\mathrm{g}}$ (\textit{middle column}) and temporal evolution of the radial profile of the dust surface density $\Sigma_{\mathrm{d}}$ (\textit{right column}) for the planet masses $M_p=0.06$, $3.74$ and $210.8\,M_{\oplus}$ (\textit{top}, \textit{middle} and \textit{bottom row}, respectively).}
    \label{fig:dust}
\end{figure*}

\begin{figure}
	\includegraphics[scale=0.395]{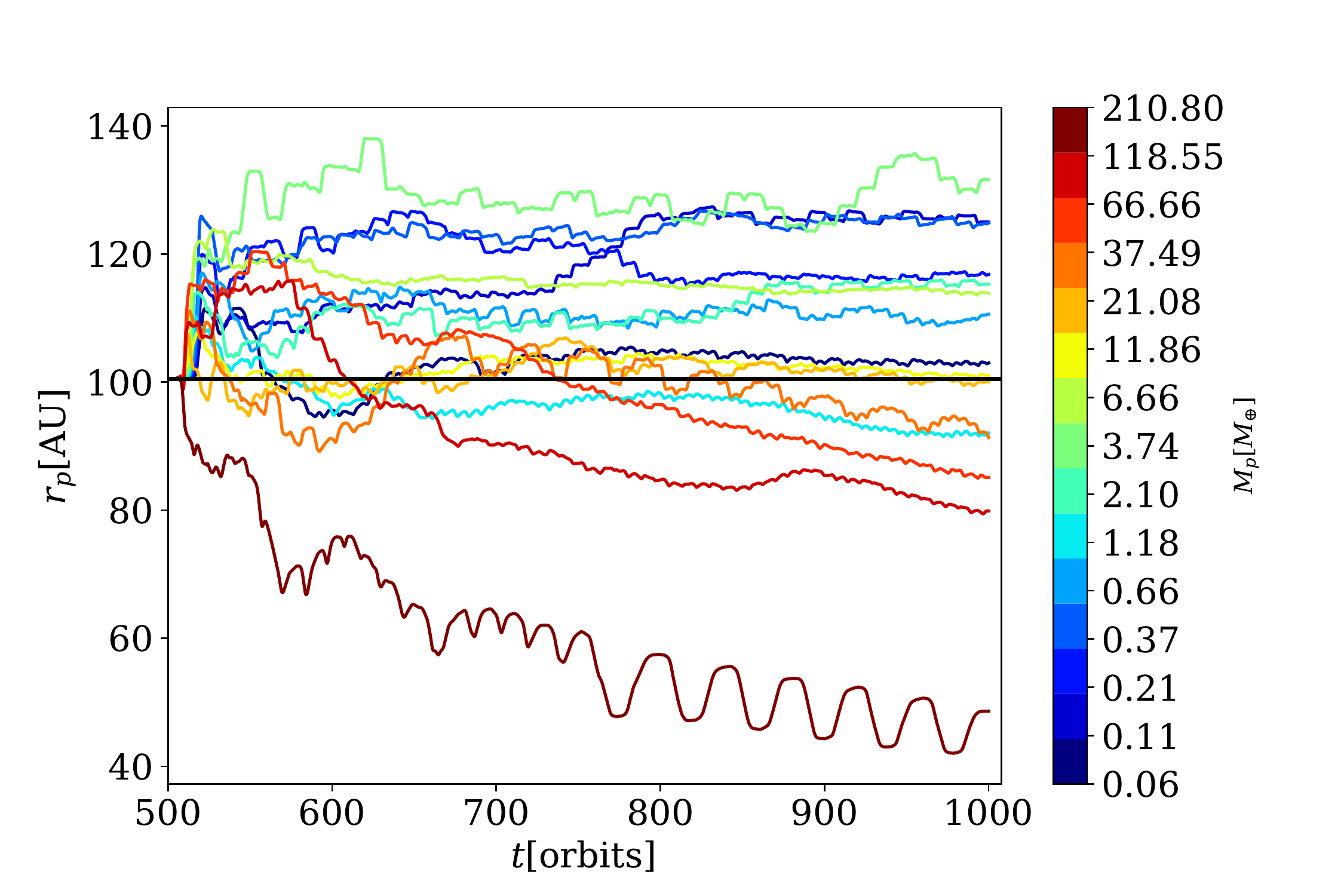}
    \caption{Temporal evolution of the semimajor axis of the planets with logarithmically spaced masses $M_p$ $\in$ $[0.06-210.80]M_{\oplus}$ in the presence of spreading pressure bumps. The horizontal black line shows the initial orbit of the planets.}
    \label{fig:Semi}
\end{figure}
\begin{figure*}
	\includegraphics[scale=0.7]{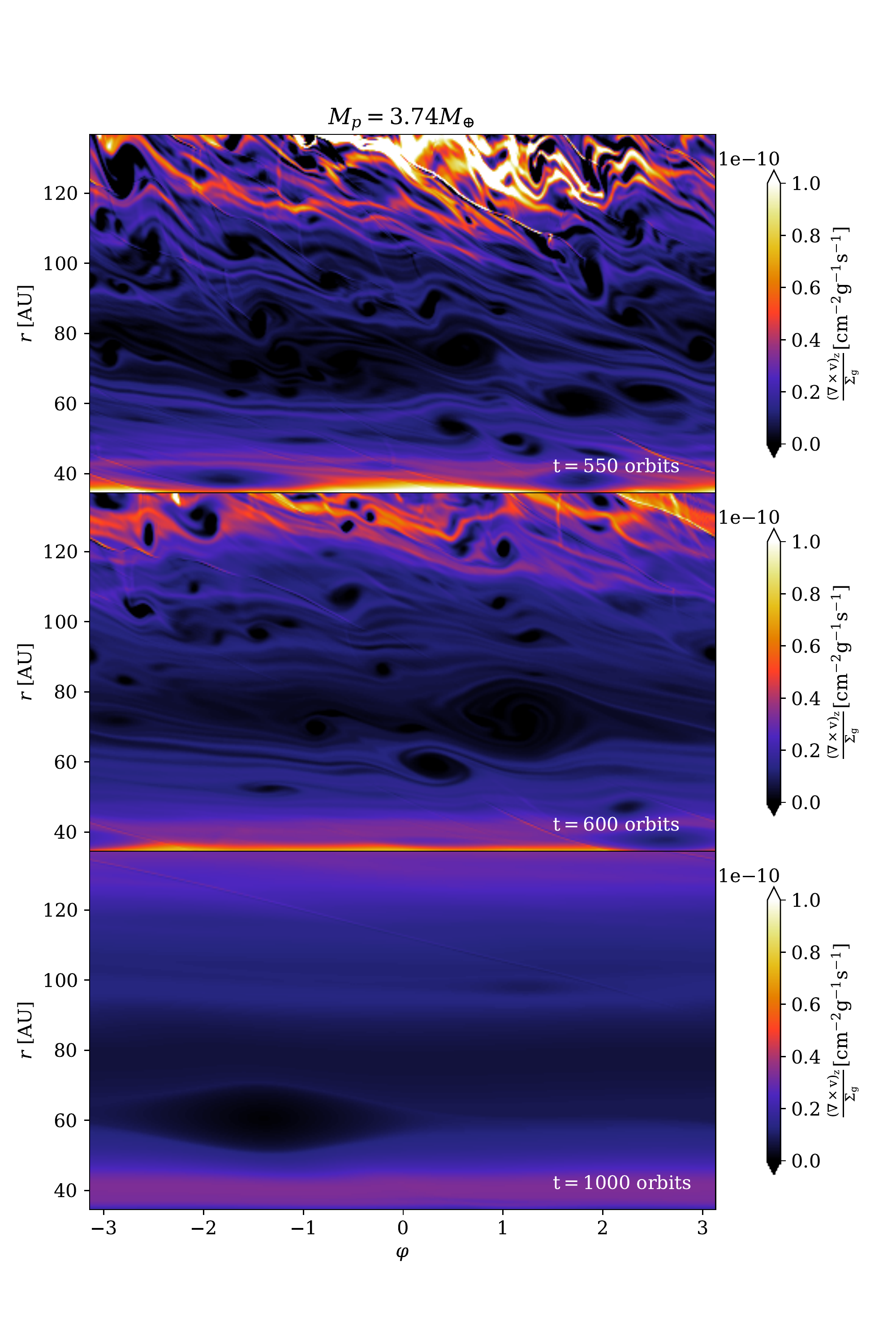}
    \caption{Temporal evolution of the vortensity considering a planet with $M_{\mathrm{p}}=3.74M_{\oplus}$. We can see that vortices are spawned throughout the disc due to the RWI. These vortices interact with the planet and modify its migration.}
    \label{fig:vort}
\end{figure*}

\subsection{Dust}

In this study, dust is treated in the pressureless fluid approximation and it is allowed to exchange its momentum with gas. The continuity and momentum equations for dust read
\begin{equation}
\frac{\partial\Sigma_{\mathrm{d}}}{\partial t}+\nabla\cdot(\Sigma_{\mathrm{d}}\mathbf{v}+\mathbf{j})=0,
\end{equation}
and
\begin{equation}
\Sigma_{\mathrm{d}}\frac{D\mathbf{v}}{Dt}=-\Sigma_{\mathrm{d}}\nabla\Phi+\Sigma_{\mathrm{d}}\mathbf{f}_{\mathrm{d}},
\end{equation}
respectively, where $\mathbf{v}$ is the dust velocity and $\mathbf{j}$ is the mass flux due to diffusion of dust particles \citep{Morfil1984}.

Dust grains are considered to have uniform sizes $a$
that determine the level of aerodynamic coupling, which is typically characterized by the dimensionless stopping time referred to as the Stokes number
\begin{equation}
\mathrm{St}=\frac{\pi}{2}\frac{a\rho_{\mathrm{m}}}{\Sigma_{\mathrm{g}}},
\label{eq:Stokes_number}
\end{equation}
where $\rho_{\mathrm{m}}=1$ $\textrm{g\,cm}^{-3}$ is the intrinsic material density of dust. Equation (\ref{eq:Stokes_number}) is valid when dust grains are smaller than the mean free path of the gas molecules and the drag operates in so-called Epstein regime. Considering that the aerodynamic interaction between gas and dust is linear in the relative velocity, the acceleration term takes the form
\begin{equation}
\mathbf{f}_{\mathrm{d}}=\frac{\Omega_{\mathrm{K}}}{\mathrm{St}}(\mathbf{u}-\mathbf{v}).
\end{equation}
In our study, we consider one dust species with $a=5\,\mu\mathrm{m}$, which translates into a Stokes number of $St=3.74\times10^{-5}$ at $r=r_0$. Therefore, gas and dust remain tightly coupled which allows us to extrapolate the gas surface density profile directly from the dust density profile reported in \citet{2021ApJ...912..164D} without taking into account any viscosity transitions in the gas disc.

\subsection{Planet}
 We use a reference frame centred on the star and rotating with the angular frequency $\Omega_{\mathrm{frame}}=\sqrt{GM_{\star}/r_{0}}$. The gravitational potential is given by
\begin{equation}
\Phi = -\frac{GM_{\star}}{r} -\frac{GM_{p}}{\sqrt{r_p^2+\epsilon^2}}+\frac{GM_pr\cos\phi}{r_p^2}+G\int_S\frac{dm(\mathbf{r}')}{r'^3}\textbf{r}\cdot\textbf{r}'
\label{eq:potP}
\end{equation}
where $M_\star=2M_\odot$ is the mass of the central star, $M_p\in[0.06$--$210.80]\,M_{\oplus}$ is the planet mass, and $r_p\equiv\abs{\mathbf{r}-\mathbf{r}_p}$ is the distance from the planet. Furthermore, $\epsilon$ is the softening length used to model the effects of nonzero vertical thickness of the disc. We use $\epsilon=0.6H_p$, where $H_p$ is the disk scale height at $r=r_p$. The third and fourth terms in Eq.~(\ref{eq:potP}) are the indirect terms due to the planet and the gravitational force of the disc, respectively, which appear because the reference frame is non-inertial and offset from the system's barycentre. The integration in the indirect term of the disc is performed over the disc surface.

We point out that the interval of $M_{p}$ considered in our study satisfies the condition
$M_p/M_\star<h^3$ \citep[][]{2006ApJ...652..730M} for the flow linearity in the planet's vicinity
because we set the uniform aspect ratio $h=0.1$ \citep[in the same way as in][]{2013A&A...557A.133D,Pinte2018,2021arXiv211107416C}. With this choice of $h$,
planets remain in the type I migration regime because
we avoid formation of a circumplanetary disc and also an increase in the corotation torque that could give rise to type III migration.
Therefore, we do not exclude any material from the inner part of the planetary Hill sphere when calculating
the gravitational torque since such a cutoff is not justified for planets without circumplanetary
discs.

\begin{figure*}
	\includegraphics[scale=0.7]{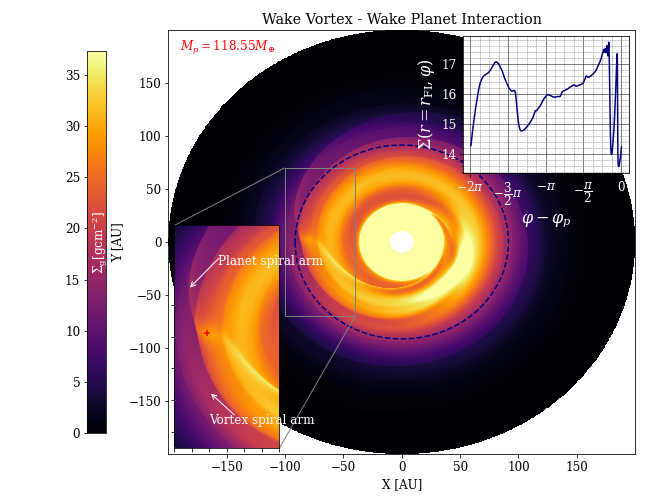}
    \caption{2D map of the gas surface density $\Sigma_{\mathrm{g}}$ in Cartesian coordinates at $t=1000$ orbits. In the zoomed region (\textit{lower left corner}), we show the interaction between the outer spiral planet-induced arm and the outer spiral vortex-induced arm for the planet with $M_p=118.55\,M_{\oplus}$. The position of the planet is indicated by the red cross. 
    The azimuthal profile of the surface density (\textit{upper right corner}) is measured along
    the dashed blue circle at $r=r_{\mathrm{FI}}=91.4\,\mathrm{au}$ and indicates the interaction
    of two shock fronts.}
    \label{fig:wake-w}
\end{figure*}
\begin{figure}
	\includegraphics[scale=0.57]{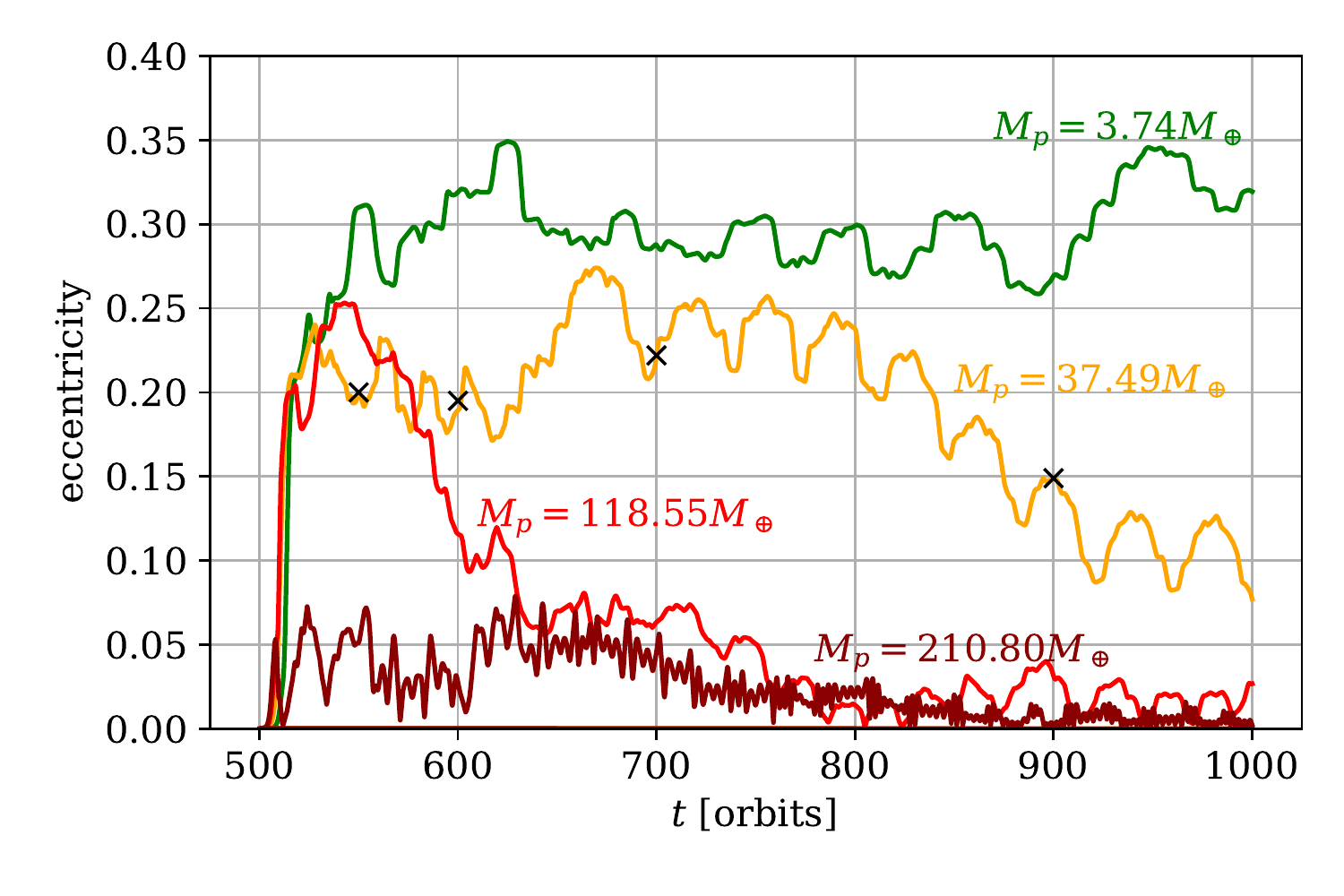}
    \caption{Eccentricity growth of four selected planets $M_{\mathrm{p}}=3.74$ (green curve), $37.49$ (orange curve), $118.55$, (light red curve)
    and $210.8\,M_{\oplus}$ (dark red curve) evolving in the presence
    of spreading pressure bumps. The black cross markers on the orange curve indicate when $\varphi_{\mathrm{vortex}}-\varphi_p=0$.}
    \label{fig:ecc}
\end{figure}
\subsection{Code and numerical setup}
We use the multifluid finite difference code FARGO3D \citep{Benitez2016,Benitez2019},
which uses the fast orbital advection algorithm of \citet{Masset2000} to significantly increase the integration timestep in numerical studies of astrophysical discs. For the mesh resolution, we use $N_{r}=1238$ cells covering 
$0.4-8.0r_0$ in the radial direction and $N_{\phi}=1024$ cells covering the whole azimuth. Therefore, cells at the reference radius $r_0=25$ au have a roughly square shape
and dimensions $\Delta r=\Delta \phi=6.13\times10^{-3}r_0$.
Note that the corotation torque is sufficiently resolved with the given mesh spacing for planets with $M_p\geq0.66\,M_\oplus$. We have compared the cell sizes to the half-width of the horseshoe region $x_s$ for $M_p=0.66\,M_\oplus$ and $M_p=210.80\,M_\oplus$ (the highest considered mass). 
The relation $x_s=1.16r_p(M_p/hM_\star)^{1/2}$ is appropriate for the former while $x_s\approx2.45r_p(M_p/3M_\star)^{1/3}$ is appropriate for the latter, using the results given in
\citet{2006ApJ...652..730M}. Setting $r_p=4.02r_{0}$, the resolution per the half-width of the horseshoe region 
is $\sim3$ and $\sim76$ cells for $M_p=0.66$ and $210.8\,M_{\oplus}$, respectively.
Although the resolution of $x_s$ for $M_p=0.66\,M_{\oplus}$ might appear relatively poor, 
we verified that doubling the resolution does not qualitatively change the outcome of our simulations.
Additionally, we point out that a similar resolution was successfully applied in \citet{Ataiee_etal2014}
and the corotation torque was well recovered.

For the initial dust surface density, we use a radial profile motivated by the solution of \citet{2021ApJ...912..164D} \citep[see also][]{2021A&A...647A.174R} that fit the observations of HD 163296. The profile at $t=0$ can be written as
\begin{equation}
\begin{split}
\Sigma_d&=\Sigma_{\textrm{inner}}\exp{[-(r/r_0)^5]}\\&+\Sigma_{\textrm{b1}}\exp{[-(r-r_{\textrm{b1}})^2/2\mathscr{W}^2_{\textrm{b1}}]}\\&+\Sigma_{\textrm{b2}}\exp{[-(r-r_{\textrm{b2}})^2/2\mathscr{W}^2_{\textrm{b2}}],}
\end{split}
\label{eq:dust_profile}
\end{equation}
where the dust surface density in the inner disc region is $\Sigma_{\textrm{inner}}=20.66$ $\textrm{g\,cm}^{-2}$ at $r=r_0$.
We consider two bumps positioned at  $r=r_{\textrm{b1}}$ and $r=r_{\textrm{b2}}$ where $\Sigma_{\textrm{b1}}=1.26$ and $\Sigma_{\textrm{b2}}=0.58$ $\textrm{g\,cm}^{-2}$, respectively. The radial widths of dust rings are $\mathscr{W}_1=4.0$ and $\mathscr{W}_2=3.9$ $\textrm{au}$ for the respective pressure bumps.

Adopting the dust-to-gas ratio $1\%$ of \citet{2021A&A...647A.174R} and considering that the gas and dust remain tightly coupled, we can infer the initial gas surface density $\Sigma_{g}$ by scaling Eq.~(\ref{eq:dust_profile}).
The initial gas distribution is shown in
Fig. (\ref{fig:initial}).
In our model, the peaks of $\Sigma_{\mathrm{g}}$ trace the pressure bumps (via Eq.~\ref{eq:eos}) and similar to several previous studies
 \citep{Lovelaceetal1999,Lietal2000,Lietal2001,Meheutetal2010,Takietal2016}, the pressure bumps are initially supported by the azimuthal velocity profile
 \begin{equation}
v_{\phi}=\sqrt{\frac{GM}{r}+\frac{r}{\rho}\nabla_rP}.
 \label{eq:Vx}
\end{equation}

We use a low-viscosity regime and adopt the $\alpha$-viscosity description \citep{Shakura1973} with $\alpha=10^{-4}$ \citep[][and  references therein]{2015MNRAS.454.1117S}. To avoid spurious wave reflections at the radial boundaries of our computational mesh, we use damping boundary conditions for the density and velocity perturbations as in \citet{Val2006}. The damping is implemented similarly to \citet[][see their Eq. 7]{2016ApJ...826...13B}, with the widths of the inner
and outer damping rings equal to $1.60$ and $27.64$ au, respectively. We use a damping time-scale equal to 1/20th of the local orbital period at the edge of each damping ring.

\section{RESULTS}
\label{sec:results}
\subsection{Single planet migrating near spreading pressure bumps}
Before a planet is inserted, the disc is allowed to evolve over 500 orbital periods (see Fig. \ref{fig:dust}) to test that the pressure bumps remain stable if left unperturbed and to clearly distinguish the disturbance
of the pressure bumps in the presence of the planet (see Sect.~\ref{sub:disc_structures}).
In Fig. \ref{fig:dust}, curves at $t=500$ orbits reveal that the pressure bumps slightly decrease in amplitude and spread radially because the disc viscously evolves. Apart from viscous evolution, the stability of pressure bumps is ensured.

Having verified that the pressure bumps remain in rotational equilibrium for at least 500 orbital periods, we insert the planet at $r=4.02r_0$ (where the maximum of the outer pressure bump is located) with null eccentricity $e=0$. To avoid numerical noise, we gently introduce the planet, allowing it to reach its full mass during the first five orbits.
The evolution of surface densities in Fig.~\ref{fig:dust} shows
that pressure bumps in gas and the corresponding rings of dust spread and merge. The resulting single pressure bump is of greater radial width but lower amplitude (Fig.~\ref{fig:dust}, middle column). After 1000 orbits, both gas and dust reach a quasi-stationary state and their evolution slows down.

Because the dust in our simulations has $\mathrm{St}\ll1$ and because we neglect the dust growth and settling,
the surface density profile of the dust has the same shape as $\Sigma_{\mathrm{g}}$ at $t=1000$ orbits
(compare the middle and right columns in Fig. \ref{fig:dust}).
The tight aerodynamic coupling also implies that the settling parameter defined in \citet[][]{2021ApJ...912..164D} becomes $f_{\mathrm{set}}=1$ and therefore the aspect ratio of the dust $h_\mathrm{d}$ is equal to the aspect ratio of the gas $h$. For our simplified model, this choice of $f_{\mathrm{set}}$ is a good compromise since at $t=0$ it corresponds fairly well to the observed HD163296 image \citep[see Fig. 5 in][]{2021ApJ...912..164D}. Additionally, 
with our choice of $St\ll1$ we discard the possibility that the pressure bumps become unstable
due to a gradual accumulation of dust grains, which might occur for larger grain sizes \citep[$St\simeq 1$;][]{Takietal2016,2017EP&S...69...50O,2020DPS....5220501C}.

The redistribution of $\Sigma_{\mathrm{g}}$ and $\Sigma_{\mathrm{d}}$ has an impact on the migration of embedded planets, as shown in Fig. \ref{fig:Semi}. Shortly after the two initial pressure bumps start to spread and merge, most of the planets exhibit an outward jump in the semimajor axis, only the planet with $M_{\mathrm{p}}=210.8\,M_{\oplus}$ starts to migrate inwards. Subsequently, the semimajor axes do not evolve in an orderly fashion. In general, it appears that less massive planets settle outwards from their initial location (thus their migration is predominantly outward) while more massive planets end up on orbits inwards from their initial location (thus their migration is predominantly inward). However, there are exceptions to this behavior, e.g. the case of $M_{\mathrm{p}}=1.18\,M_{\oplus}$, and moreover, many planets exhibit episodes of substantial inward and outward excursions in the disc. The planet with $M_{\mathrm{p}}=210\,M_{\oplus}$ settles near $\simeq50$ $\mathrm{au}$, close to the position of the newly formed pressure bump.

\subsection{Vortex formation and vortex-planet interactions}

We found that the redistribution of pressure bumps is accompanied by the occurrence of vortices in all our simulations. This is due to the fact that radial variations in the density, which accompany the spreading of pressure bumps, inevitably spawn small vortices in various places of the protoplanetary disc \citep{Lietal2000}.
As illustrated in Fig. \ref{fig:vort} for a planet of $M_p=3.74\,M_\oplus$, a large number of small vortices have already formed in the disc at $t=550$ orbits ($50$ orbits after planet insertion).
These vortices quickly decay or merge to form a single large-scale vortex that can be seen at
$r\approx65$ au, appearing already at $600$ orbits and surviving to at least $1000$ orbits.
The large vortex remains trapped at the new pressure maximum.
Our results concerning the formation of vortices by the RWI mechanism are in agreement with \citet{2010ApJ...725..146P}.

Once a large vortex has formed, it generates spiral density waves similar to those generated by an embedded planet. These density waves form two spiral arms, one internal (leading) and the other external (trailing) with respect to the position of the vortex. Due to the size of the vortex, its spiral waves become strong, their density contrast being comparable or even more pronounced compared to planet-induced spiral waves. Therefore, vortex-induced spiral waves can affect migrating planets because they periodically change the density distribution and thus the torque felt by the planet. They either pass directly through the vicinity of the migrating planet or they interfere with the planet-induced spiral waves. To distinguish this new result from previous studies, we classify two possible forms of planet-vortex interactions:
\begin{enumerate}
  \item[1] \textit{Closer Interaction}. Planet within the corotation region of the vortex formation. 
  \item[2] \textit{Faraway Interaction}. Vortex spiral waves - planet spiral waves interaction.
\end{enumerate}
\begin{figure*}
	\includegraphics[width=\textwidth]{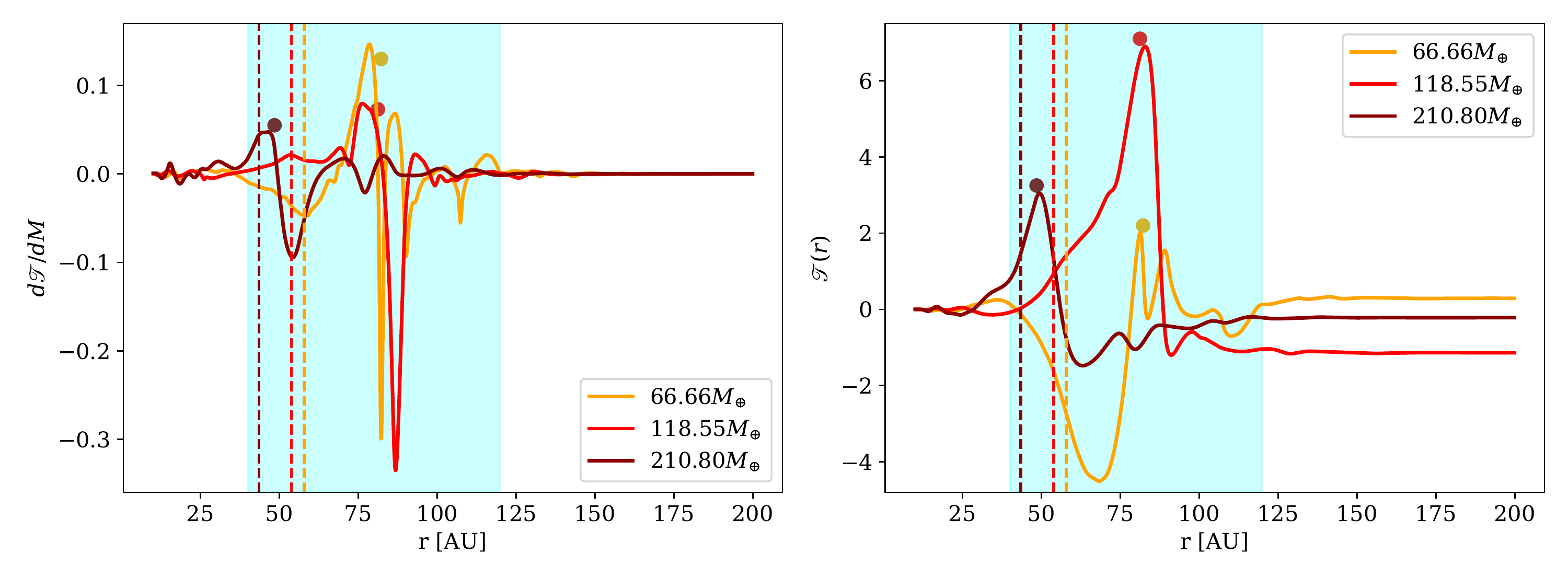}
    \caption{\textit{Left}. Torque per unit disc mass on the three planets of masses $M_p=66.66$, $M_p=118.55$ and $M_p=210.80M_{\oplus}$ as a function of the radius (in units of $r_0$). \textit{Right}. Cumulative torque acting on the planets. The $d\mathscr{T}/dM$ and $\mathscr{T}$ quantities are normalized by $\Sigma r_p^4\Omega^2\mu^2h^{-3}$. Circular points mark the location
    of the respective planets and the colored dashed vertical lines represent the positions of the vortices in each case.}
    \label{fig:Tcm}
\end{figure*}
\begin{figure}
	\includegraphics[scale=0.57]{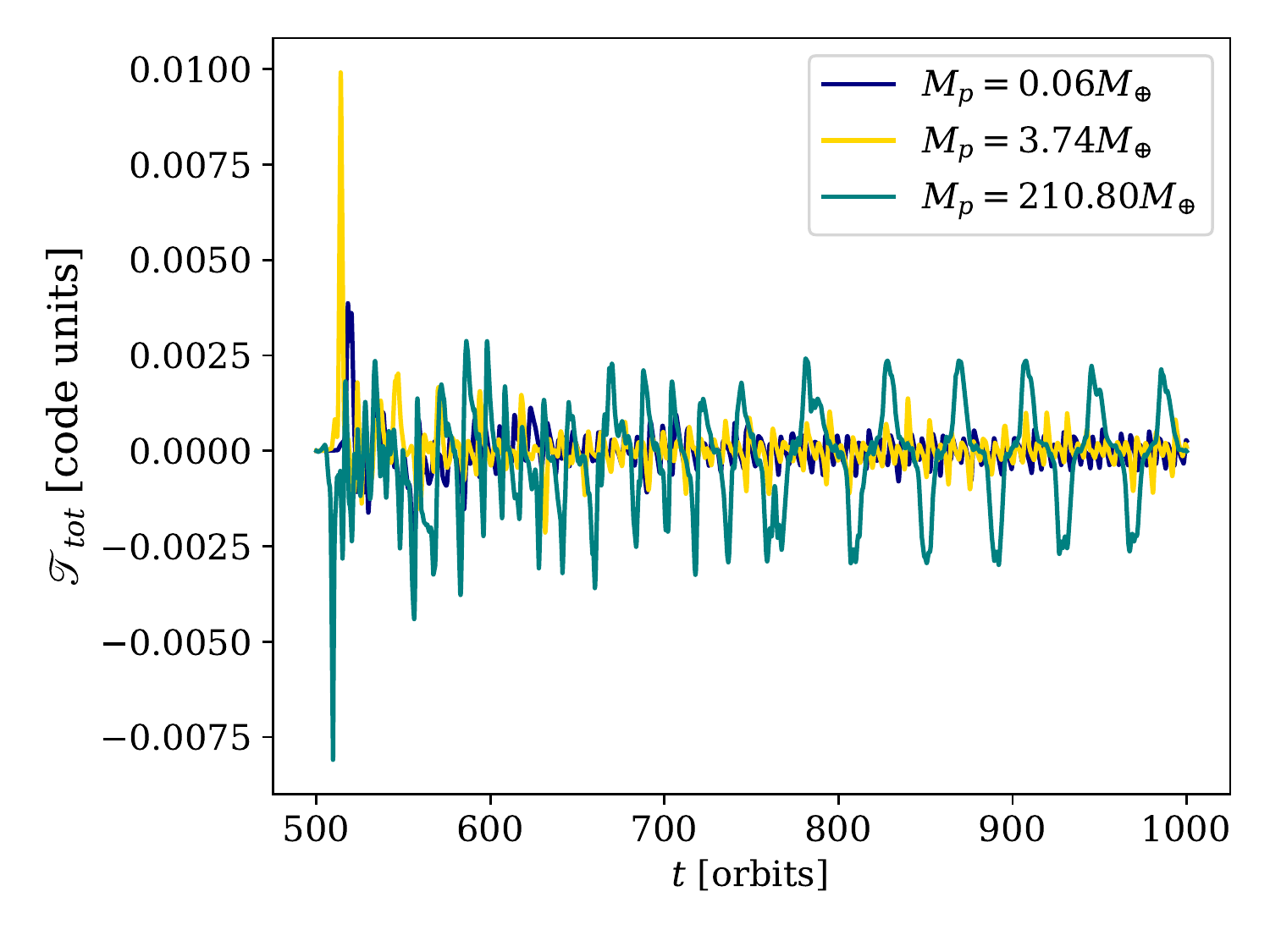}
    \caption{Total torque $\mathscr{T}_{tot}$ versus time for three planets. The oscillations are caused by the interaction of the spiral arms of the planet with the vortex spiral arms (see Appendix \ref{app:vortex_interaction} for details).}
    \label{fig:Tt}
\end{figure}
The first form of interaction between a planet and a vortex has been previously studied in detail for low-mass planets ($M_p\leq 10^{-6}M_\oplus$) in \citet{Ataiee_etal2014}. In our study, the prevailing form of interaction between the planet and the vortex is the second, Faraway Interaction. As we mentioned above, the vortex remains at a considerable distance ($r\simeq65\,\mathrm{au}$) from the planets ($r=80$--$130$ AU in most cases). Fig. \ref{fig:wake-w} shows a snapshot of the gas surface density for the $M_p=118.55M_\oplus$ planet that we can use to demonstrate the principle of the Faraway Interaction.
One can see that the outer spiral arm of the vortex interferes with the outer spiral arm of the planet.
The strength of the density perturbation is shown in the inset in the upper right corner of Fig.~\ref{fig:wake-w}. The inset measures the surface density along the whole azimuth at 
$r_{\mathrm{FI}}=91.4$ au (along the blue dashed line in Fig.~\ref{fig:wake-w}).
There is a sharp N-shaped profile at $\varphi\simeq\varphi_p$ and a peak between $\varphi=-7/4\pi$ and $\varphi=-3/2\pi$ that are driven by the passage of the outer arm of the vortex. In Appendix, we confirm the mechanism of Faraway Interaction using a standard model for a vortex embedded in a protoplanetary disc \citep[similar to that described in][]{2010ApJ...725..146P}, but in this case interacting with a planet of mass $M_p=3.74\,M_\oplus$.

By the vortex-planet interactions, we can explain the oscillations and stagnation of
$r_{\mathrm{p}}(t)$ (see Fig. \ref{fig:Semi}).
Moreover, we found excitation of orbital eccentricities in some cases.
In Fig.~\ref{fig:ecc}, we show the temporal evolution of the eccentricity for four planets. We measured that the eccentricity can increase up to $e=0.1$--$0.3$ for the lowest mass planets considered in this study.

Let us take a closer look at the eccentricity evolution in one specific case
$M_{\mathrm{p}}=3.74\,M_{\oplus}$ (green curve in Fig.~\ref{fig:ecc})
and its relation to the evolution of vortices in the disc, as represented by the vortensity maps in Fig.~\ref{fig:vort}.
Shortly after the planet insertion in the simulation, the eccentricity jumps
from 0 to $\simeq$$0.3$. While this is partly due to the overall redistribution of
the background disc density profile, we also see that many small-scale vortices appear
in the disc ($t=550$ orbits in Fig.~\ref{fig:vort}) and they inevitably interact with the planet. As the time progresses, the small-scale vortices gradually vanish ($t=600$ orbits in Fig.~\ref{fig:vort}) until a single large Rossby vortex remains ($t=600$--$1000$ orbits in Fig.~\ref{fig:vort}). Despite the disappearance of small-scale vortices, 
the eccentricity of the $M_{\mathrm{p}}=3.74\,M_{\oplus}$ planet is maintained in the time interval $t=600$--$1000$ orbits. Such an evolution differs from
the standard eccentricity damping that is usually dominant for disc-embedded planets
(the damping is exhibited e.g. by $M_{\mathrm{p}}=118.55$ and $210.8\,M_{\oplus}$ in Fig.~\ref{fig:ecc}). We infer that the eccentricity damping is prevented by the interaction of the $M_{\mathrm{p}}=3.74\,M_{\oplus}$ planet with the large vortex, which forms early ($t=600$ orbits) and survives until $t=1000$ orbits. We point out that the interaction is substantial despite the large orbital separation $\Delta r \simeq 70 \, \mathrm{au}$ between the planet and the vortex.

In the case of $M_p=37.49M_\oplus$, it can be seen that the eccentricity is maintained above the value of 0.2 up to $t\approx850$ orbits, then it decreases and remains at a value of $e\approx0.1$.
To rule out the possibility that some changes of the eccentricity are driven by conjunctions
of the planet and the vortex, we put several crosses on the orange curve in Fig.~\ref{fig:ecc}
that mark the times when the vortex and the planet coincide in their azimuthal positions, that is, $\varphi_{\mathrm{vortex}}-\varphi_p=0$. We find that there is no pattern between the 
eccentricity evolution and the points where a vortex-planet passage (in azimuth) occurs. Therefore,
the Faraway Interaction dominates.

Next, we analyze the torque density distribution and the temporal evolution of the total torque exerted on the planet by the whole disc (gas and dust combined).
The torque density per unit disc mass as a function of radius $d\mathscr{T}/dM$ is given as \citep[][]{DAngelo_etal2010}
\begin{equation}
\mathscr{T}=2\pi\int^{\infty}_0\frac{d\mathscr{T}}{dM}\Sigma(r)rdr
\label{eq:Tcm}
\end{equation}
where $\Sigma(r)$ is the azimuthally averaged gas surface density and $\mathscr{T}$ is the cumulative torque. In Fig. \ref{fig:Tcm}, we show $d\mathscr{T}/dM$ and $\mathscr{T}$ as functions of the radius (in units of $r_0$) at $t=1000$ orbits for planets with masses $M_p=66.66$, $M_p=118.55$ and $M_p=210.80\,M_{\oplus}$. At this time, the planets are located at $r_p=3.39r_0$, $r_p=3.25r_0$ and $r_p=1.94r_0$ (marked with dots in Fig. \ref{fig:Tcm}), respectively. We find that the radial torque distribution shows strong variations in the external part of the disc beyond the planetary orbits, which can lead to outward migration. We think that the well-defined positive/negative peaks in the radial torque distribution are due to changes in density near of the planet produced by the interaction of the spiral arms generated by a vortex (see Appendix \ref{app:vortex_interaction}) because when the outer spiral arm of the vortex interferes with the inner (outer) spiral arm of the planet, it can causes an increase (decrease) in density in the vicinity of the planet, since except for the planet of greater mass, the vortices in each of our models are located at a great distance of the planets. On the other hand, we can also see that the cumulative torque has the largest contribution within the radial extension of the pressure bump (colored area in Fig. \ref{fig:Tcm}). In the particular case of the planet with mass $M_p=66.66\,M_{\oplus}$, the cumulative torque is negative for $r<r_p$ within the radial extension of the bump, while it is positive beyond the orbital radius of the planet (see the right panel in Fig. \ref{fig:Tcm}). This behavior in the radial profile of the cumulative torque can be explained because the formation of the large vortex takes a little longer, and a slightly smaller vortex survives in the innermost region of the disc, which also generates spiral waves. So the spiral arms of the planet interact simultaneously with the outer spiral arms of these two vortices. The pitch angle of each of the outer spiral arms of the vortices is similar to that of the planetary inner spiral arm and thus the latter does not suffer a great disturbance. However, there is a disturbance in density because the merge process of the two vortices continues, which gives a negative contribution to the cumulative torque. Conversely, the outer spiral arm of the planet frequently suffers strong interference with the outer spiral arms of each vortex and it is therefore weakened. The latter results in a much slower migration of the planet (see Fig. \ref{fig:Semi}).

Fig. \ref{fig:Tt} shows the temporal evolution of the total torque for the planets with $M_p=0.06$, $M_p=3.74$ and $M_p=210.80M_{\oplus}$. The torque is oscillating, which corresponds to the oscillations
of the semi-major axes. The amplitude of oscillations is greater during the first 550 orbits,
which may be due to the spread of the initial pressure bumps or due to the appearance of multiple vortices.
However, after this transitional phase, the oscillations become more regular and we can attribute
this behaviour of the total torque to the interaction between the vortex-induced and planet-induced
spiral arms. In other words, the Faraway Interaction of the planets with a large vortex is the main mechanism driving the migration of planets with $M_{p}\leq118.55\,M_{\oplus}$. 
In the case of the $M_p=210.80\,M_\oplus$ planet, the vortex-induced spiral waves are no longer
strong enough to fully dominate the torque and the planet thus migrates inwards until it reaches
the proximity of the large vortex and eventually undergoes the Closer Interaction between 750 and 1000 orbits. In fact, this last case gives us the possibility of differentiating the effect generated in the total torque on the planet by the Faraway Interaction from the Closer Interaction, since the curve of the total torque exhibits asymmetric oscillations when the Faraway Interaction dominates whereas the oscillations become symmetric once the Closer Interaction, which has a well defined periodic behavior, takes over. During this last time interval, the planet exhibits classical oscillations in its semi-major axis, as known for planets interacting directly with a vortex \citep[see][]{Ataiee_etal2014}.
To summarize, the majority of planets in our simulations interact with the outer
spiral arm of the vortex rather than with the core of the vortex itself. The case of $M_{\mathrm{p}}=210.8\,M_{\oplus}$ is an exception. In the respective simulation, the planet migrates all the way to the direct proximity of the vortex.

\subsection{Stability analysis}
\label{sub:disc_structures}

Formation of vortices in our simulations (Fig.~\ref{fig:vort})
is necessarily a result of a hydrodynamic instability.
In previous studies, it has been found that a maximum of unstable pressure in the protoplanetary disc\footnote{Here, the maximum of unstable pressure is considered in the context of violation of the Solberg-Hoiland criterion $\kappa^2(r)+N^2(N)\geq0$, with $\kappa$ and $N$ being the radial epicyclic and radial Brunt-V\"ais\"al\"a frequencies, respectively.} can activate the so-called Rossby Wave Instability (RWI) which gives rise to vortices in the gas disc. These vortices have been studied extensively in the context of redistribution of the angular momentum in the disc \citep[see ][and references therein]{Lietal2001}. 

An immediate way to know if the vortex is caused by the RWI is to show that the function (which includes the vorticity of the equilibrium flow)
\begin{equation}
\begin{split}
\mathcal{L}&=\frac{\Sigma_g\Omega}{\kappa^2}\left(P\Sigma_g^{-\gamma}\right)^{\frac{2}{\gamma}}\\ &=\frac{\Sigma_g}{(2\nabla\times \mathbf{v})_z}\left(P\Sigma_g^{-\gamma}\right)^{\frac{2}{\gamma}}
\label{eq:Lvort}
\end{split}
\end{equation}
has an extremum. For barotropic discs, equation (\ref{eq:Lvort}) can be written as
\begin{equation}
\mathcal{L}=\frac{\Sigma_g}{(2\nabla\times \mathbf{v})_z}.
\label{eq:Lbar}
\end{equation}

\begin{figure}
	\includegraphics[scale=0.55]{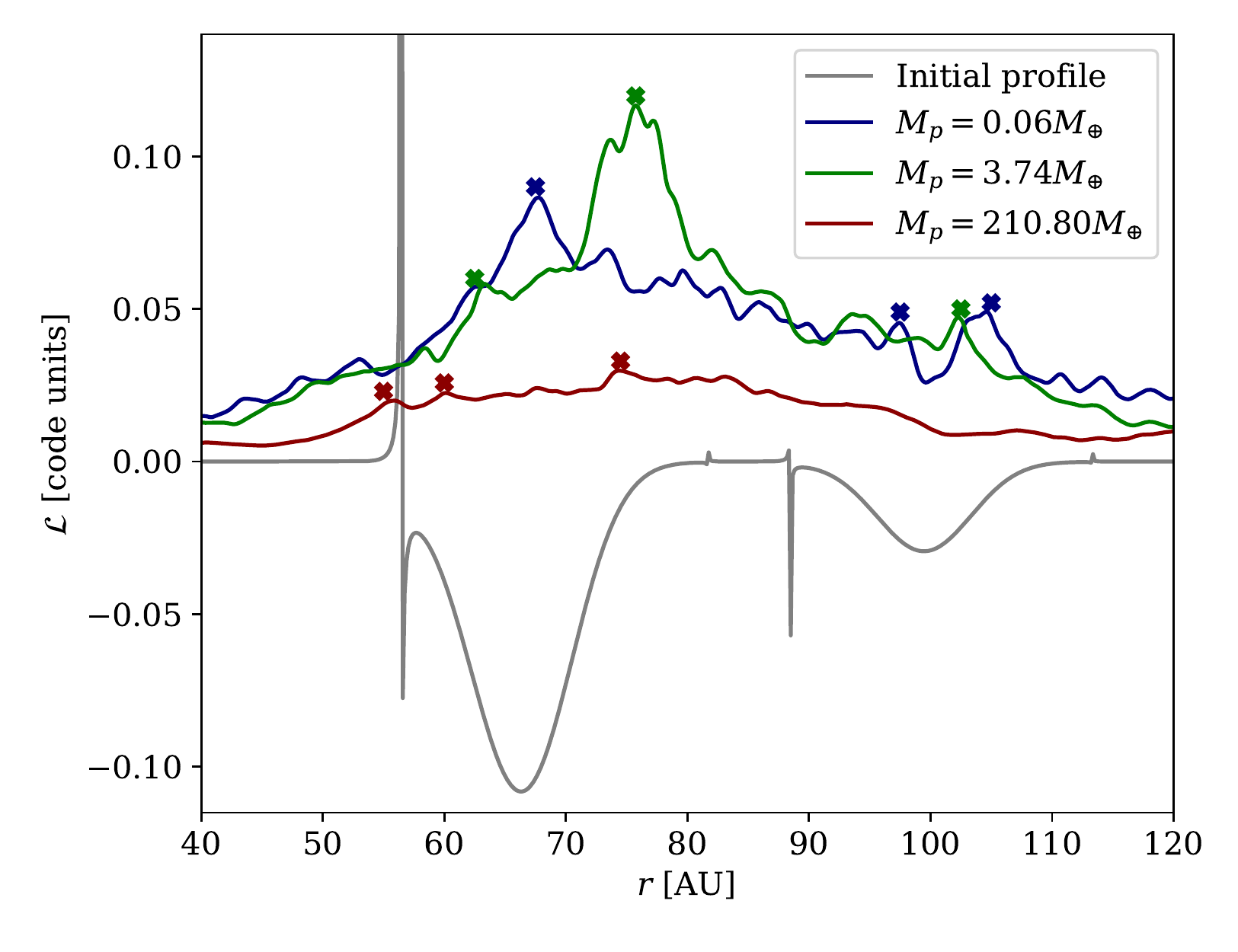}
    \caption{Radial profile of $\mathcal{L}$ (azimuthally averaged) given in Equation (\ref{eq:Lbar}) for three different planets at $t=550$ orbits (50 orbits after the planet was inserted in the disc). The colored crosses mark the positions of three small vortices in each case. The positions of the vortices coincide with the inflection points of $\mathcal{L}$. Note that in the case of the initial profile of $\mathcal{L}$, the peaks that appear are due to advective mixing of material.}
    \label{fig:L}
\end{figure}

For all planetary masses from our parametric space, the radial profile for $\mathcal{L}$ calculated from equation (\ref{eq:Lbar}) suggests vortex formation due to the RWI. Figure (\ref{fig:L}) shows the radial profile of $\mathcal{L}$ azimuthally averaged at $t=550$ orbits for $M_{\mathrm{p}}=0.06$, $3.74$ and $210.8M_{\oplus}$. Since several small vortices formed before $t=550$ (see the upper panel of Fig. \ref{fig:vort}), we only include the positions of three such vortices (shown in Fig.~\ref{fig:L} by crosses). It is clear that the vortices form at positions where the radial profile of $\mathcal{L}$
has inflection points. In addition, Fig.~\ref{fig:L} shows the profile of $\mathcal{L}$ before the planets are inserted and we find that the vortensity exhibits features at the pressure bumps
related to the well-known mixing of material and homogenization of the vortensity distribution \citep[][]{2007LPI....38.2289W,2009MNRAS.394.2283P,2021MNRAS.508.2329C}.

Additionally, one might ask whether the pressure bumps are indeed stable to vortex formation before
a planet is inserted or whether our demonstration of the bump stability in Fig.~\ref{fig:dust} is merely a lucky outcome of perfectly symmetric initial conditions. To answer the question, we restarted the 
simulation interval $t=0$--$500$ orbits and introduced a random noise in the radial velocity
component of the gas with an amplitude of $5\%$ of the local sound speed. We found only a very little
change compared to surface density profiles shown in Figure \ref{fig:dust} and the bumps again remained
stable to vortex generation.

We think that the spread of the pressure bumps and the generation of vortices in all our studied models is mainly because the greatest change in the vortensity, $\Delta \zeta$, is due to the shocks produced by the planet's wakes. \citet{2021MNRAS.508.2329C} showed that
\begin{equation}
\Delta\zeta\propto\left(\frac{M_p}{M_{\mathrm{th}}}\right)^3
\label{eq:delta_z}
\end{equation}
where 
$M_{\mathrm{th}}=c_s^3/\Omega G$ is the thermal mass. In addition, they found that low-mass planets with $M_p\ll M_{\mathrm{th}}$ can produce asymmetric jumps in the vortensity inwards/outwards from their orbit in places when the disc scale height increases. We mention that a study of the critical planetary mass that is necessary to destabilize a pressure bump and generate a vortex is outside the scope of this paper
and it will be explored in detail in a forthcoming work (Chrenko \& Chametla, in preparation).

\begin{figure*}
	\includegraphics[scale=0.47]{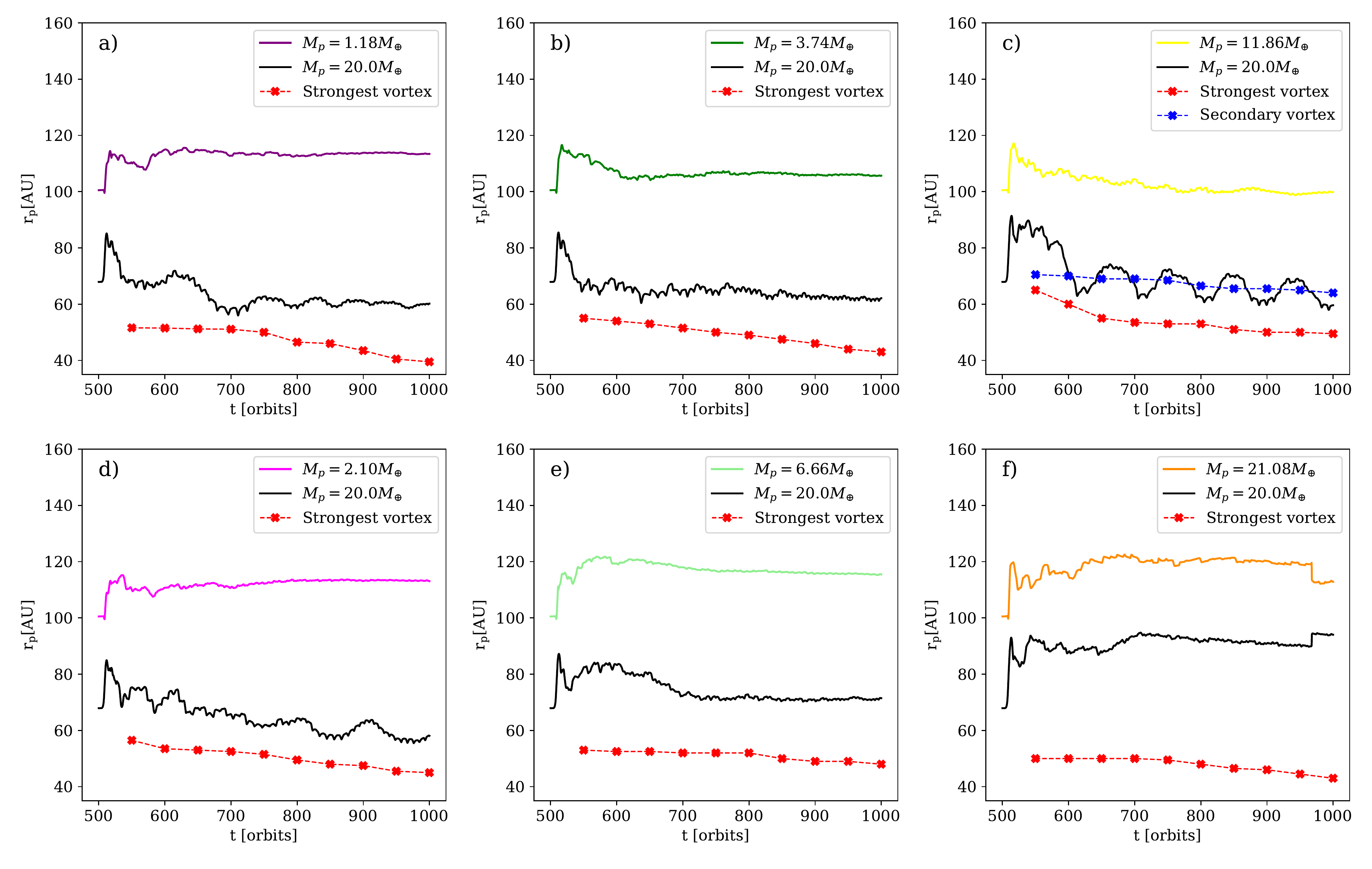}
    \caption{Temporal evolution of the semimajor axes of planet pairs and the vortex positions in our simulations. In all cases (a--f) it can be seen that the strongest vortex can only interact with the planets via the \textit{Faraway Interaction} since the orbits of the planets never cross the position of the vortex. In case c), the position of a secondary vortex is shown to demonstrate the effects produced by the \textit{closer interaction}. }
    \label{fig:Semi2}
\end{figure*}
\begin{figure}
	\includegraphics[scale=0.375]{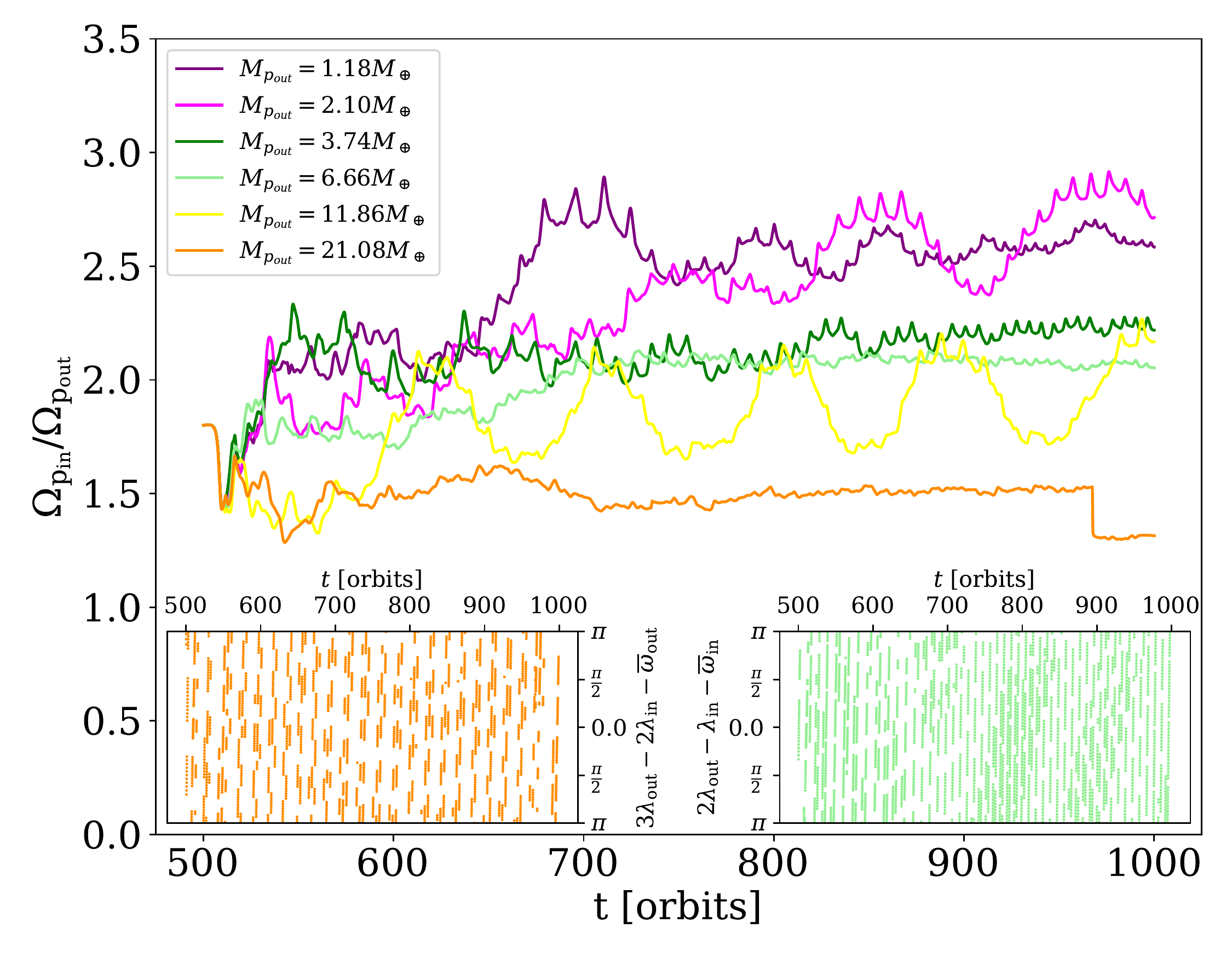}
    \caption{Temporal evolution of the ratio of the angular velocities, $\Omega_{\mathrm{p_{in}}}$ and $\Omega_{\mathrm{p_{out}}}$, which correspond to the planets initially located at the inner and outer pressure bump, respectively. The boxes show the resonant angles $\phi'=2\lambda_{\mathrm{out}}-\lambda_{\mathrm{in}}-\overline{\omega}_{\mathrm{in}}$ (\textit{lower right corner}) and $\phi=3\lambda_{\mathrm{out}}-\lambda_{\mathrm{in}}-\overline{\omega}_{\mathrm{out}}$ (\textit{lower left corner}) for the pairs with the outer planet mass $M_{p_{\mathrm{out}}}=6.66M_\oplus$ and $M_{p_{\mathrm{out}}}=21.08M_\oplus$, respectively. Here $\lambda_{\mathrm{in}}$ and $\lambda_{\mathrm{out}}$ are the mean longitudes of the inner and outer planet. In a similar manner, $\overline{\omega}_{\mathrm{in}}$ and $\overline{\omega}_{\mathrm{out}}$ are the length of the pericenter of the inner and outer planet, respectively.}
    \label{fig:Ome}
\end{figure}
\subsection{Migration of two planets}
Here we study the configuration in which there is a planet located at the pressure maximum of each bump. The mass of the inner planet located at the internal pressure bump with the orbital radius $R_{p_\mathrm{in}}=69.8\,\mathrm{au}$ is fixed to $M_{p_\mathrm{in}}=20\,M_\oplus$. The mass of the second planet, initially located at $R_{p_{\mathrm{out}}}=100.5\,\mathrm{au}$, is varied logarithmically in the interval $[0.66,21.08]\,M_\oplus$. Once the planets are inserted in the disc and allowed to migrate, the response of the disc is similar to our simulations with a single planet: the two pressure bumps are destabilized, the disc density is redistributed and the formation of multiple vortices occurs again. However, we found that the number of large vortices that form depends on the mass of the outer planet (because the mass of the inner planet remains fixed). In addition, the time it takes for these large vortices to merge also depends on the mass of the outer planet as we discuss below. For $M_{p_{\mathrm{out}}}\in [0.66,6.66]M_\oplus$ we find the formation of three large vortices, for $M_{p_{\mathrm{out}}}\leq 11.86M_\oplus$ four large vortices and for $M_{p_{\mathrm{out}}}\leq 21.08M_\oplus$ we find formation of at least six large well-defined vortices. In all cases, vortex formation takes place in less than 50 orbits. 

In Fig. \ref{fig:Semi2}, we show the temporal evolution of the semi-major axes of six pairs of planets migrating through the two pressure bumps (labels $\mathrm{a-f}$ in Fig. \ref{fig:Semi2} ordered from the lowest to the largest mass of the outer planet, respectively). Due to formation of multiple large vortices, the orbital dynamics of each planet (inner / outer) is governed by the two types of planet-vortex interactions described above (Closer and Faraway Interactions). The contribution of each of these planet-vortex interactions depends on the time it takes for the vortices to merge and also on the relative distance between vortices and planets. For cases $\mathrm{a)}$ and $\mathrm{d)}$, the three initial vortices exist until $t\simeq700$ orbits, a merging event occurs afterwards and two vortices survive up to $1000$ orbits. The inner planet suffers both types of interactions with the vortices. In cases $\mathrm{b)}$ and $\mathrm{e)}$, the evolution is similar but at $t\simeq800$, two remaining vortices merge into a single vortex. 
 Therefore, the inner planet undergoes both types of interactions for $t\lesssim800$ orbits 
 but only the Faraway Interaction for $t\gtrsim800$ orbits. For the case $\mathrm{c)}$, four vortices are formed at first and they merge to form two vortices in less than 150 orbits. These two remaining vortices interact in both ways with the inner planet up to $t =1000$ orbits (see panel c) in Fig. \ref{fig:Semi2}). We think that the Closer Interaction generates the oscillating behavior of the semi major axis of the inner planet and that the Faraway Interaction keeps the planet in a stagnant migration on average. Finally, for the case $\mathrm{f)}$, we found that the six initial vortices merge into two vortices in less than 200 orbits, these two vortices survive and both planets interact with them through the Faraway Interaction.
 We point out that for the outer planet, the Faraway Interaction leads to a stagnant migration in all cases $\mathrm{a)-f)}$ (see Fig. \ref{fig:Semi2}).
 
On the other hand, Fig. \ref{fig:Semi2} might suggest that the pairs of planets could become trapped in a Mean Motion Resonance ($\mathrm{MMR}$) during their migration. In order to analyze this possible behavior, we calculated the temporal evolution of the ratio of the angular velocities for each planet pair (see Fig. \ref{fig:Ome}). In addition, we calculated several resonant angles for each planet pair, taking into account the strongest first-order resonances such as $2:1$ and $3:2$, as well as some others like $5:2$. We found that the resonant angles do not exhibit a librating pattern and so the planets cannot be firmly trapped in $\mathrm{MMRs}$. As a demonstration, we include in Fig.~\ref{fig:Ome}
the evolution of two critical angles $\phi'=2\lambda_{\mathrm{out}}-\lambda_{\mathrm{in}}-\overline{\omega}_{\mathrm{in}}$ (box in the lower right corner) and $\phi=3\lambda_{\mathrm{out}}-\lambda_{\mathrm{in}}-\overline{\omega}_{\mathrm{out}}$ (box in the lower left corner) for the cases $e)$ and $f)$ of Fig. \ref{fig:Semi2}. Clearly, $\phi'$ and $\phi$ oscillate. Therefore, the pairs of planets in cases $e)$ and $f)$ are not trapped in the resonances $2:1$ and $3:2$, respectively, although the ratios of the planetary angular velocities suggest a proximity to these resonances. However, we do not rule out the possibility that a resonant trapping may occur at $t>1000$ orbits.

Even if there is no resonant trapping, the results of this section clearly indicate that due to the Faraway Interaction, the orbital behavior of both planets is very similar, with outward migration episodes or stagnant migration. Therefore, we think that the planet-vortex interaction mechanism can also govern the planetary migration of multiple systems in protoplanetary discs where vortex formation takes place.

\section{Conclusions}
\label{sec:conclusions}
In this study, we analyzed the effect of two pressure bumps previously formed in a gas-dust disc on the migration of low- to intermediate-mass planets, adopting a density profile of gas and dust as extrapolated from the recent studies of the HD163294 disc \citep{2021ApJ...912..164D,2021A&A...647A.174R}. We introduce a planet at the maximum pressure of the outer bump (located at $r_{\mathrm{b2}}=100.5$ au) and we find that both pressure bumps spread and form a single pressure bump which is radially wider but has a much lower amplitude compared to the initial bumps. The redistribution of the gas and dust density during the spreading of pressure bumps leads to vortex formation due to the Rossby Wave Instability (RWI). At first, several mini-vortices interact with the planet but they gradually vanish or merge into larger vortices.
In most of our simulations, a typical outcome
is an appearance of a large vortex in the inner disc at a considerable distance from the planet.
This vortex generates density waves that resemble the spiral waves emitted by the planet. The spiral arms of the vortex interact through interference with the spiral arms of the planet that is migrating in the disc. This interaction substantially modifies the total torque exerted on the planet and causes a slow and/or stagnant migration. We defined this planet-vortex interaction as \textit{Faraway Interaction} since it only occurs through the interaction of the vortex- and planet-induced spiral waves  while the vortex core and the planet themselves remain distant. In other words, there is no direct interaction of the planet with the vortex core.

For the case of two planets migrating from each pressure bump, the migration is also stalled or exhibits oscillations.
This behavior depends mainly on the time it takes for the large vortices to form. Additionally, 
we found that the formation of multiple large vortices is possible in this case. Typically,
the inner planet (initially located at the first pressure maximum) interacts with the core of a vortex as well as through the Faraway Interaction. For the outer planet (located initially at the second pressure maximum), the migration is governed only by the Faraway Interaction with multiple vortices.
We ruled out any efficient trapping of the planet pairs in mean-motion resonances during their migration (at least for the chosen set of parameters and simulation time spans).

We conclude that the Faraway Interaction is a very powerful interaction mechanism between a planet and a vortex that can slow down the classical inward Type I migration or even revert it in some cases.

\section*{Acknowledgements}
We are grateful to the referee for her/his constructive and careful report which significantly improved the quality of the manuscript. This work was supported by the Czech Science Foundation (grant 21-23067M). The work of O.C. was supported by Charles University Research program (No. UNCE/SCI/023). Computational resources were available thanks to a Marcos Moshinsky Chair and to UNAM's PAPIIT grant BG101620. The authors are grateful to F. Javier S\'anchez Salcedo for useful discussions and for his advice in many aspects.

\section*{Data Availability}

The FARGO3D code is available from \href{http://fargo.in2p3.fr}{http://fargo.in2p3.fr}. The input files for generating our multifluid 2D hydrodynamical simulations will be shared on reasonable request to the corresponding author.




\bibliographystyle{mnras}
\bibliography{mnras_template} 




\appendix
\section{Simplified model of the planet-vortex interaction}
\label{app:vortex_interaction}
\begin{figure*}
	\includegraphics[scale=0.5]{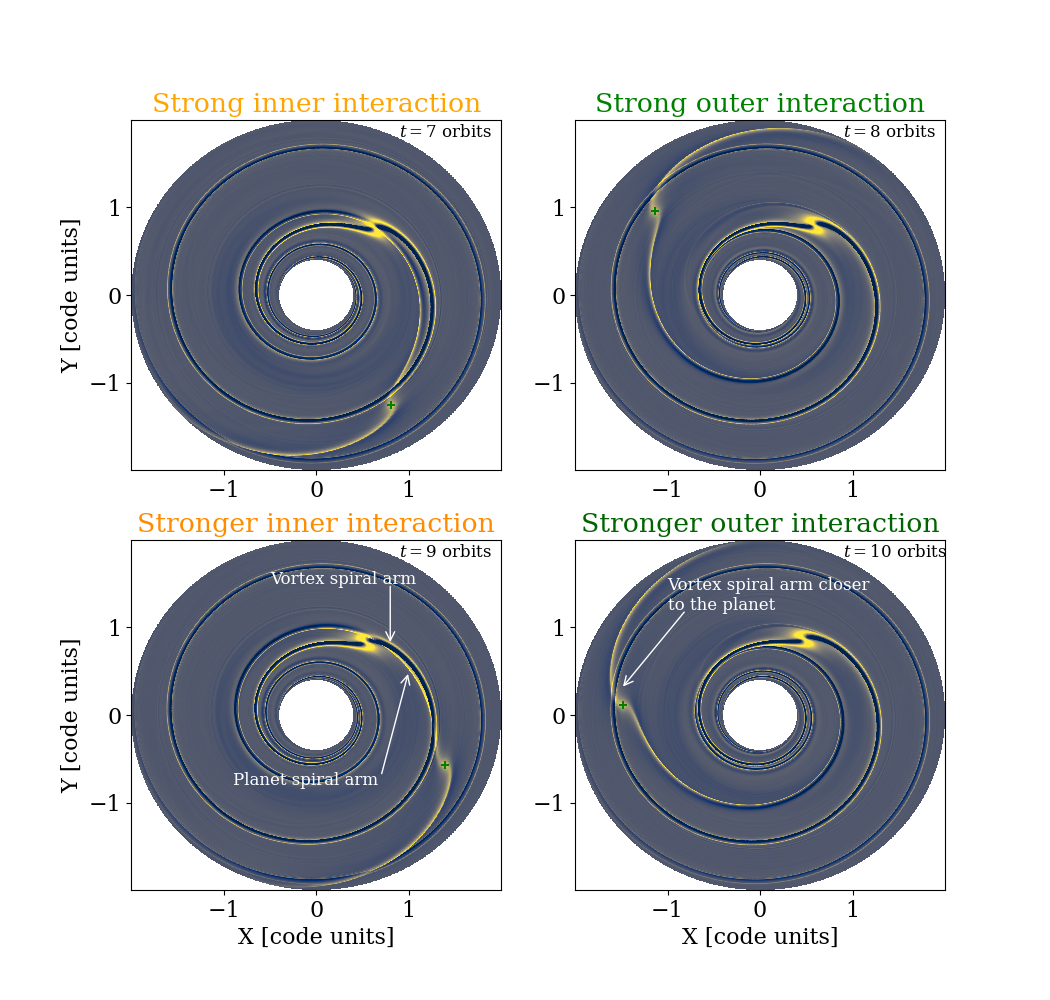}
    \caption{2D maps of the gas surface density for four consecutive time periods where possible faraway interactions between the vortex and the planet are shown. It is interesting to note that the outer spiral arm of the vortex generates strong interference in both spiral arms of the planet very close to the position of the planet (marked with the green plus symbol).}
    \label{fig:Vortex10}
\end{figure*}

The interaction between planets and vortices has been studied in the context of a closer interaction of the vortex core with a planet \citep[][]{2011MNRAS.415.1426L,Ataiee_etal2014}. However, because the vortex can generate spiral density waves \citep[][]{2009MNRAS.397...52H,2009MNRAS.397...64H,2010A&A...513A..60L,2010ApJ...725..146P,2012MNRAS.426.3211L}, a distant interaction between the planet and the vortex is possible (which we call the \textit{Faraway Interaction}) and it is governed mainly by the shocks between the spiral arms of the vortex and the spiral arms generated by the planet. Since this interaction shows a periodic behavior, the migration of the planet is substantially modified. To show the effects of the Faraway Interaction, we performed a 2D hydrodynamic simulation of a vortex induced in a smooth power-law gas disc interacting with a planet located far from the vortex core.

Our setup is similar to the fiducial case of \citet{2010ApJ...725..146P} with some slight modifications. We consider an isothermal disc with a surface density profile given as power law $\Sigma=\Sigma_0(r/r_0)^{-s}$.  Here $r_0$ is the initial position of the vortex, $\Sigma_0$ is arbitrary and we use the $\alpha$ viscosity of $10^{-4}$. To produce the vortex in the gas disc, we introduce an initial velocity perturbation of $0.5 c_s$ over a circular region of radius $H_0/2$ located at $(r,\varphi)=(1,0)$. We use $H_0=0.05r_0$ and $s=0$ to avoid vortex migration
(the angular momentum of the vortex scales as $H_0^2$ and vortex migration is therefore faster in thicker discs; similarly, vortex asymmetries make vortex migration strongly dependent on $s$). Therefore, we guarantee that the planet-vortex interaction is only facilitated by spiral waves emitted by both. We modeled a full disc with a radial extension from $0.4\,r_0$ to $2.0\,r_0$ using a mesh resolution of $(N_r,N_\varphi)=(900,1200)$ zones. The planet was introduced into the gas disc at $r=1.5\,r_0$ with a mass of $M_p=3.74\,M_\oplus$.

\begin{figure}
	\includegraphics[scale=0.5]{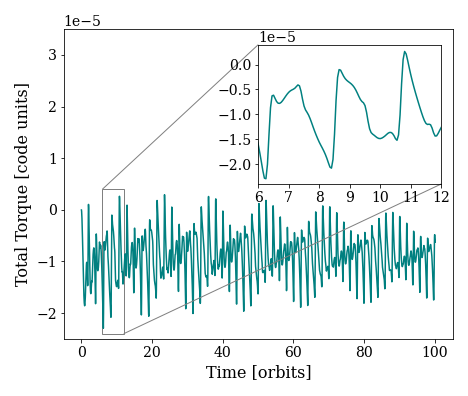}
    \caption{Temporal evolution of the total torque felt by a planet during its interactions with the outer spiral arm of a vortex. The inset shows the behavior of the total torque during orbital times discussed in Appendix A.}
    \label{fig:Tt_append}
\end{figure}

\begin{figure}
	\includegraphics[scale=0.55]{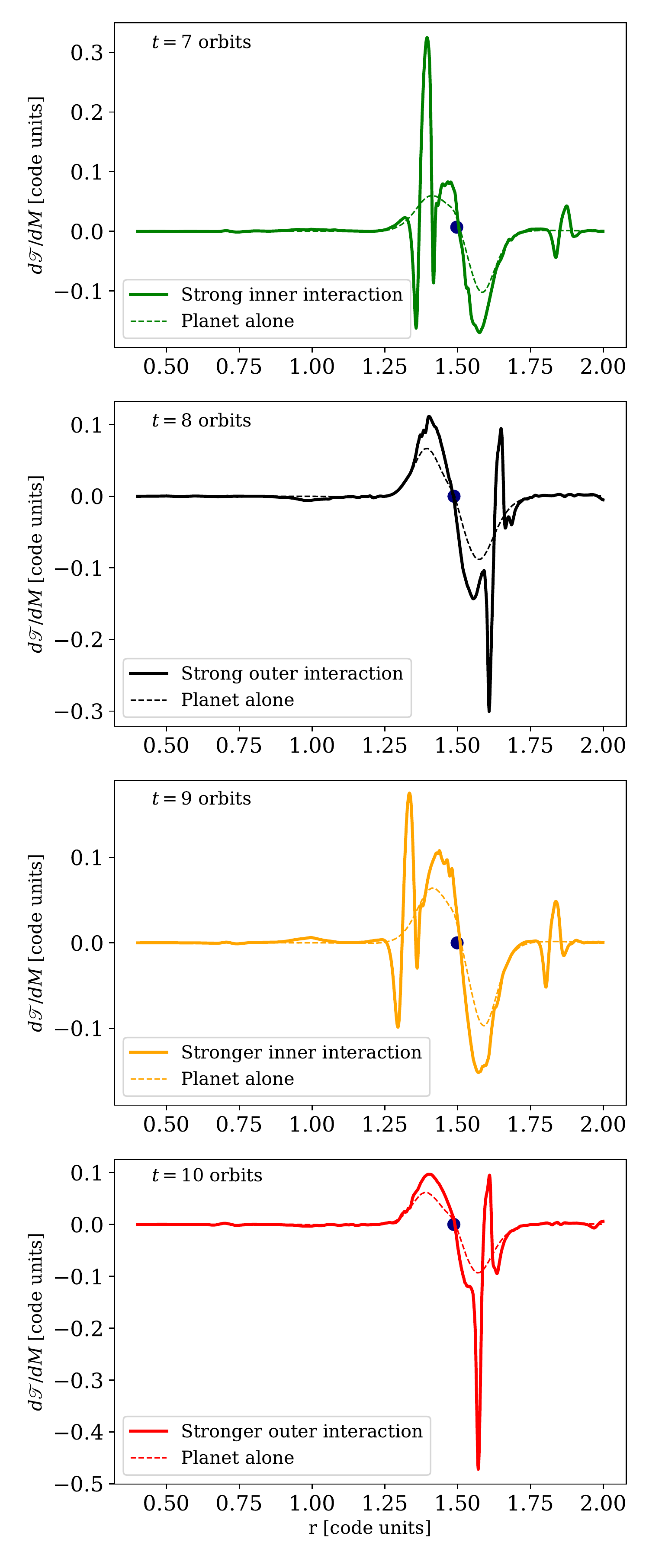}
    \caption{Torque per unit disc mass calculated at the same orbital times as in Fig. \ref{fig:Vortex10}. Circular blue points mark the location of the planet.}
     \label{fig:dT_A}
\end{figure}

In Fig. \ref{fig:Vortex10}, we show the interactions between the planet spiral arms and the vortex spiral arms at $t=7$, $8$, $9$ and $10$ orbits. Clearly, it can be seen that the outer spiral arm of the vortex interacts with the spiral arms of the planet each orbital period . At $t=7$ orbits, the outer spiral arm of the vortex interacts with the inner spiral arm of the planet and at $t=8$ orbits, the outer spiral arm of the vortex now interacts with the outer spiral arm of the planet. Therefore, the interaction of the outer vortex-induced spiral arm with the two spiral arms of the planet takes consecutive turns. These interference between the spiral arms produce a substantial change in the planet's torque which becomes oscillating because the perturbation is periodic (see Fig. \ref{fig:Tt_append}). In fact, we find that the outer spiral arm of the vortex interacts more strongly with the inner (outer) spiral arm of the planet every two orbital periods. As can be seen at $t=9$, the outer spiral arm of the vortex almost completely overlaps the inner spiral arm of the planet. At $t=10$ orbits, on the other hand, the outer spiral arm of the vortex generates a strong interference near the planet in its outer spiral arm. The inner spiral arm of the vortex contributes to a lesser extent since it interacts only with the inner spiral arm of the planet and the interaction is weak because it is very distant. 

As a final demonstration of the effects described above, we show in Fig. \ref{fig:dT_A}
the torque density per unit disc mass as a function of radius $d\mathscr{T}/dM$
measured again at $t=7$, $8$, $9$ and $10$ orbits and compared with the torque density generated
by a planet alone without any vortex. It can be clearly seen that the greatest differences 
in $d\mathscr{T}/dM$ occur at exactly the same radii where the spiral arms of the vortex interfere with the planet's vicinity and its spiral arms. We emphasize that a strong inner interaction of the vortex arms with the planetary spiral arms results in a higher positive torque than in the case of an isolated planet (see panels for $t=7$ and $t=9$ orbits in Fig. \ref{fig:dT_A}). On the other hand, an outer interaction produces a negative change in the torque at $t=8$ and $t=10$ orbits. Therefore, we think that the oscillations observed in Fig. \ref{fig:Tt_append} can be explained entirely by the influence of the spiral arms of the vortex. The most prominent peaks in the torque density profiles for $t=7$, $8$, $9$ and $10$ orbits
that alternate between positive and negative contributions to the torque
can be associated with the changes in magnitude in the zoom of Fig. \ref{fig:Tt_append}.

We conclude that the \textit{Faraway Interaction} can effectively change the total torque felt by the planet embedded in the gas disc and considerably modify planet migration.


\bsp	
\label{lastpage}
\end{document}